\documentclass[12pt]{article}

\usepackage[a4paper,total={18cm,27cm}]{geometry}

\usepackage[utf8]{inputenc}
\usepackage{amsmath}
\usepackage{amsfonts}

\usepackage{graphicx}
\usepackage{array}
\usepackage{caption}

\usepackage{xcolor}
\colorlet{shadecolor}{yellow!20}

\usepackage{authblk}

\usepackage{hyperref}

\usepackage{comment}   

\newcommand{\NI}{\vspace{0.2cm}\noindent}

\begin{document}


\title{Convergent Evolution in Algorithmic Space}


\author[1,2,3,4]{Patrick Krauss}
\author[2,3,4]{Achim Schilling}
\author[1]{Andreas Maier}
\author[2,4]{Thomas Kinfe}
\author[3]{Henri Stübner}
\author[3]{Niklas Römmelt}
\author[1]{Claus Metzner}

\affil[1]{\small Cognitive Computational Neuroscience Group, Pattern Recognition Lab, Friedrich-Alexander-University Erlangen-Nürnberg (FAU), Germany}
\affil[2]{\small Neuromodulation and Neuroprosthetics, University Hospital Mannheim, University Heidelberg, Germany}
\affil[3]{\small Neuroscience Lab, University Hospital Erlangen, Germany}
\affil[4]{\small BG Clinic Ludwigshafen, Germany}

\maketitle


\begin{abstract}
In evolutionary biology, unrelated organisms can independently evolve similar structures when exposed to similar functional demands. Here we ask whether an analogous form of convergent evolution occurs during neural network training: do networks with different random initializations develop similar internal weight structures when trained on the same task? This question is technically nontrivial because hidden neurons can be arbitrarily permuted without changing the represented function, making direct matrix comparisons misleading. We introduce a matching-based framework for comparing multilayer perceptrons in structural weight space. Hidden neurons are first coarsely aligned using permutation-invariant features and then refined by iterative Hungarian matching. After alignment, networks are compared with structural distance metrics designed to emphasize task-relevant weight patterns. Applying this approach to ensembles of small MLPs trained on MNIST, Fashion-MNIST, and KMNIST, we find that networks trained on the same task remain closer to one another than to networks trained on different tasks. Thus, task-specific training appears to guide initially random networks toward distinct regions, or attractors, in structural network space. The earliest phase of training reveals an additional and unexpected phenomenon. Classification accuracy rises rapidly before the matched structural distances show strong task-specific separation, and before the global weight distribution visibly changes. Nevertheless, individual weight entries already begin to drift in a coordinated manner. This suggests that early learning may first operate through subtle, distributed adjustments that strongly affect function while leaving coarse network morphology almost unchanged. We treat this early morphogenesis as a first glimpse of a richer dynamical process that will be investigated in future work.

\end{abstract}

\newpage

\section{Introduction}

\NI Convergent evolution is one of the clearest indications that biological form is not shaped by historical contingency alone. Unrelated organisms can independently develop similar functional solutions when they are exposed to comparable environmental constraints. Sharks and dolphins, for example, belong to fundamentally different evolutionary lineages, yet both evolved streamlined bodies that reduce drag during rapid movement through water. Likewise, birds and bats evolved powered flight independently. Their wings perform a similar function, but they are built from different anatomical structures and arose through different evolutionary pathways. Wings, eyes, streamlined bodies, and many other functional structures have thus emerged repeatedly in distant lineages when similar environmental constraints favored similar solutions \cite{conwayMorris2003lifes,losos2017improbable,petryshen2020evidence}. The usual interpretation is not that evolution follows a predetermined script, but that the space of possible solutions is highly structured: under comparable constraints, some regions of this space are more accessible, more stable, or more efficient than others.

\NI A closely related question can be asked for learning systems.  If several neural networks with identical architecture are trained independently on the same task, do they merely reach functionally equivalent but structurally unrelated parameter configurations, or do they converge toward similar internal organizations?  In other words, does a learning problem act like a constraint that guides independently initialized systems toward common regions of an abstract algorithmic space?  If so, neural-network training would provide a controlled model system for studying a form of convergent evolution outside biology: not the repeated emergence of anatomical structures, but the repeated emergence of algorithmic structures \cite{lecun2015deep,alzubaidi2021review,li2015convergentLearning}.

\NI This question touches Marr's distinction between computational, algorithmic, and implementational levels of analysis \cite{marr1982vision}.  The training task defines a computational problem.  The learned weights constitute a concrete implementation.  Between these levels lies the algorithmic organization: the internal representations and transformations by which the network solves the task.  Two trained networks may differ numerically at the implementational level while nevertheless sharing an algorithmic organization.  Detecting such similarities directly in weight space is therefore a way of asking whether the computational problem constrains the set of possible algorithmic solutions.

\NI In the broader motivation of this project, we view such recurring solution regions as possible algorithmic attractors.  The term is meant descriptively: an attractor is a region of parameter or representation space that independently initialized learning systems tend to approach when exposed to the same functional constraints.  This idea is compatible with recent discussions of convergent representations in artificial and biological intelligence, including the possibility that learning systems discover privileged representational or algorithmic structures rather than inventing arbitrary idiosyncratic ones \cite{li2015convergentLearning,morcos2018insights,csiszarik2021similarity,klabunde2025similarity,huh2024platonic}.  Ultimately, our long-term goal is to test whether such attractors can be detected across tasks, architectures, modalities, and levels of description, from low-level network parameters to high-level semantic representations.

\NI Previous work by our group provides the conceptual and methodological background for this study.  At the representational level, we introduced the Generalized Discrimination Value as a way to quantify class separability across layers of neural networks and observed reproducible layer-wise separability profiles across independently initialized models \cite{schilling2021quantifying}.  At the dynamical level, our studies of recurrent and reservoir networks investigated how random connectivity, nonlinear dynamics, attractor regimes, and structural regularities shape information processing \cite{metzner2022dynamics,metzner2025nonlinear,metzner2025organizational}.  Complementary work on successor representations and cognitive maps explored how neural systems can encode relational structure across spatial, visual, and linguistic domains \cite{stoewer2022successor}.  Finally, our analyses of recurrent and transformer-based language models showed that abstract linguistic categories and construction-level representations can emerge in trained networks without being explicitly supplied as labels \cite{krauss2025wordclass,ramezani2025argument,kissane2026verbparticle,koebl2026prediction}.  Together, these studies motivate the present focus on parameter-space convergence and algorithmic attractors.

\NI To turn this broader motivation into a concrete test, however, one must move from representation-level observations to a direct comparison of the trainable mechanisms themselves.  This is less straightforward than it may first appear.  Weight matrices provide an explicit record of the learned implementation, but their raw coordinates are not uniquely defined.  Hidden neurons within a layer can be permuted without changing the function implemented by the network, provided the corresponding incoming and outgoing weights are permuted consistently.  This permutation symmetry is increasingly recognized as central to the geometry of neural-network loss landscapes and model comparison \cite{entezari2022permutation,ainsworth2023git}.  Thus, two networks may be functionally and structurally similar while their weight matrices look unrelated because corresponding neurons occupy different index positions.  Any attempt to detect convergence in weight space must therefore first align the hidden neurons of the networks being compared.

\NI The present paper develops and applies such an alignment-based approach.  We combine permutation-invariant presorting features with iterative Hungarian matching to define matched distances between neural networks.  We then use these distances to study small multi-layer perceptrons trained independently on MNIST, Fashion-MNIST, and KMNIST.  The simplicity of these networks is intentional: before applying the framework to deep or highly structured architectures, we first ask whether convergent organization can be detected in the most transparent possible setting.

\NI The results support the existence of task-dependent structural convergence, but they also revealed a second, more unexpected aspect of learning dynamics.  While matched network distances show that networks trained on the same task become structurally more similar than networks trained on different tasks, the earliest phase of training is dominated by surprisingly small weight changes that nevertheless produce large improvements in loss and accuracy.  Thus, the beginning of learning appears less like a large-scale morphological reorganization of the network and more like a highly effective functional calibration of structures already latent in the random initialization.  This observation suggests that the landscape of simple learning problems may be more training-friendly than a naive picture of isolated optima would imply: the many symmetry-equivalent and redundant realizations of good solutions may make the parameter space effectively rich in nearby routes toward useful algorithmic organization.

\section{Methods}

\subsection{Training datasets}

We used three image-classification datasets with identical input format and identical train--test split sizes: MNIST, Fashion-MNIST, and KMNIST.  Each dataset consists of grayscale images of size $28 \times 28$ pixels, with ten output classes, 60,000 training examples, and 10,000 test examples.  MNIST contains handwritten digits, Fashion-MNIST contains clothing categories, and KMNIST contains Japanese Kuzushiji characters.  This choice allowed us to compare different classification tasks while keeping the input dimensionality, number of classes, and dataset size fixed.

Images were converted to tensors and normalized separately for each dataset using dataset-specific mean and standard deviation values.  For MNIST we used mean 0.1307 and standard deviation 0.3081, for Fashion-MNIST mean 0.2860 and standard deviation 0.3530, and for KMNIST mean 0.1918 and standard deviation 0.3483.  No data augmentation was applied.  All datasets were accessed through the standard \texttt{torchvision} dataset interface and cached locally in the project data directory.

\subsection{Multi-layer perceptrons and training procedure}

All main experiments used small fully connected multi-layer perceptrons with architecture
\begin{equation}
    784 \to 64 \to 10 .
\end{equation}
The 784 input units correspond to the flattened $28 \times 28$ image pixels.  The hidden layer contained 64 neurons with a ReLU nonlinearity, and the output layer contained ten logits, one for each class.  Bias terms were initialized to zero and kept fixed at zero throughout training, so that all structural comparisons were based only on the weight matrices.  In the notation used below, each trained network is therefore represented by two weight matrices, $W_0$ of size $784 \times 64$ and $W_1$ of size $64 \times 10$.

For each of the three tasks, we trained three independently initialized networks, giving nine networks in total.  The networks were labeled M1--M3 for MNIST, F1--F3 for Fashion-MNIST, and K1--K3 for KMNIST.  Different random seeds were used for all nine networks, affecting both the initial weights and the shuffled order of mini-batches.  Training used mini-batches of size 256 and cross-entropy loss.  For the uniformly sampled training trajectories used in the main distance-versus-episode analysis, we used stochastic gradient descent with learning rate 0.1 and momentum 0.9.

To obtain a moderately fine temporal resolution while keeping the notation simple, we define one training episode as half of one full pass through the training set.  Thus, $E=0$ denotes the random initialization before training, $E=1$ denotes the state after half an epoch, $E=2$ after one full epoch, and so on.  The main trajectories were saved from $E=0$ to $E=20$, corresponding to ten full training epochs.  At each saved episode, the weight matrices and the test accuracy were stored.

For the early-training analysis, we performed an additional run with the same architecture, datasets, labels, and random seeds, but with logarithmically spaced checkpoints in optimizer-update time.  This log-resolved run used Adam optimization with learning rate $10^{-3}$.  Its purpose was not to replace the uniformly sampled main trajectories, but to resolve the very beginning of training more finely than the half-epoch sampling.  The horizontal coordinate was converted back into the same episode scale by
\begin{equation}
    E = 2\,\frac{s}{S},
\end{equation}
where $s$ is the number of mini-batch updates already performed and $S$ is the number of mini-batch updates in one full epoch.  Thus, for example, a value such as $E=0.06$ corresponds to only a few mini-batch updates, long before the first half-epoch checkpoint.  This log-resolved run was used only for analyses that specifically required early temporal detail.

\subsection{The neuron permutation problem}

Feed-forward neural networks contain a simple but important symmetry: hidden neurons can be permuted without changing the function represented by the network, provided that the same permutation is applied consistently to all incoming and outgoing weights of the affected layer.  For a hidden layer $h$, represented by the columns of $W_h$ and the rows of $W_{h+1}$, a neuron permutation therefore corresponds to a simultaneous column permutation of $W_h$ and row permutation of $W_{h+1}$.  In the special case of a single square weight matrix, the analogous transformation is a simultaneous row and column permutation,
\begin{equation}
    B = P^{T} A P ,
\end{equation}
where $P$ is a permutation matrix.  Thus, the relevant object is not an independent row assignment or an independent column assignment, but one shared permutation that acts on two adjacent index sets at the same time.

This is why the standard Hungarian algorithm alone does not solve the full neuron permutation problem.  The Hungarian algorithm solves a linear assignment problem for a given cost matrix.  It can match rows to rows, columns to columns, or neuron profiles to neuron profiles, but it does not by itself enforce the global consistency constraints induced by hidden neurons in a multilayer network.  In particular, permuting one hidden layer changes two neighboring matrices simultaneously, and in networks with more than one hidden layer, the optimal permutation of one hidden layer also changes the effective comparison problem for the neighboring hidden layers.  The network alignment problem is therefore a coupled matching problem.

Our solution separates the problem into two stages.  First, we optionally perform a cheap presorting step that produces a rough but useful initial alignment.  This step is based on permutation invariant features (PIFs), i.e. summary features of each hidden neuron that do not depend on the arbitrary ordering of the other hidden neurons.  The purpose of this step is not to compute the final distance, but to move the networks into a comparable coordinate system from which local Hungarian steps can improve the alignment.  Second, we perform iterative Hungarian matching across the hidden layers.  This iterative step requires a local matrix distance $d(A,B)$, because each Hungarian cost matrix is constructed from distances between incoming--outgoing neuron profiles.  The structural distance metrics used in this work are defined below; details of the presorting and iterative matching procedures are given in the next subsections.

\subsection{PIF-based presorting}

For each hidden neuron, we construct a low-dimensional feature vector from its incoming and outgoing weights.  Consider hidden neuron $i$ in layer $h$.  Its incoming weight vector is the $i$-th column of $W_h$,
\begin{equation}
    u_i^{(h)} = W_h[:,i],
\end{equation}
and its outgoing weight vector is the $i$-th row of $W_{h+1}$,
\begin{equation}
    v_i^{(h)} = W_{h+1}[i,:].
\end{equation}
The PIF vector used here contains four scalar summaries of the incoming weights and four analogous summaries of the outgoing weights.  For the incoming vector $u_i^{(h)}$, we use its signed total input,
\begin{equation}
    \phi_{1,i}^{(h)} = \sum_k u_{ki}^{(h)},
\end{equation}
its number of nonzero incoming connections,
\begin{equation}
    \phi_{2,i}^{(h)} = \#\{k:u_{ki}^{(h)} \neq 0\},
\end{equation}
its number of positive incoming connections,
\begin{equation}
    \phi_{3,i}^{(h)} = \#\{k:u_{ki}^{(h)} > 0\},
\end{equation}
and the Euclidean norm of its incoming weight vector,
\begin{equation}
    \phi_{4,i}^{(h)} = \|u_i^{(h)}\|_2.
\end{equation}
For the outgoing vector $v_i^{(h)}$, we analogously use its signed total output,
\begin{equation}
    \phi_{5,i}^{(h)} = \sum_k v_{ik}^{(h)},
\end{equation}
its number of nonzero outgoing connections,
\begin{equation}
    \phi_{6,i}^{(h)} = \#\{k:v_{ik}^{(h)} \neq 0\},
\end{equation}
its number of positive outgoing connections,
\begin{equation}
    \phi_{7,i}^{(h)} = \#\{k:v_{ik}^{(h)} > 0\},
\end{equation}
and the Euclidean norm of its outgoing weight vector,
\begin{equation}
    \phi_{8,i}^{(h)} = \|v_i^{(h)}\|_2.
\end{equation}
Together, these eight quantities form the PIF vector
\begin{equation}
    \phi_i^{(h)}
    =
    \left(
    \phi_{1,i}^{(h)},\ldots,\phi_{8,i}^{(h)}
    \right).
\end{equation}
For the dense MLPs used in this study, the nonzero counts are usually constant within a layer, but they are retained because they make the definition applicable to sparse networks as well.

To presort one hidden layer, we compute the PIF vectors for all neurons in the corresponding hidden layer of the two networks.  Before matching, each feature dimension is standardized using the pooled mean and standard deviation across both networks.  We then compute a Euclidean cost matrix between PIF vectors and solve the resulting assignment problem with the Hungarian algorithm.  The resulting permutation is applied to the second network by permuting the corresponding columns of $W_h$ and rows of $W_{h+1}$.  This is repeated independently for all hidden layers.

PIF-based presorting is intentionally heuristic.  It is not used as the final metric and it is not intended to detect all functionally equivalent permutations.  Its role is to create a coarse initial alignment that removes obvious ordering differences and thereby helps the later iterative matching avoid poor local configurations.

\subsection{Iterative Hungarian matching}

After optional PIF-based presorting, the networks are refined by iterative Hungarian matching.  We represent a network as a list of weight matrices
\begin{equation}
    \mathcal{W} = (W_0,W_1,\ldots,W_{L-1}),
\end{equation}
where $W_h$ has shape $n_h \times n_{h+1}$.  Input and output coordinates are treated as fixed, while the hidden layers may be permuted.

For hidden neuron $i$ in layer $h$, we define its full local profile as the concatenation of incoming and outgoing weights,
\begin{equation}
    q_i^{(h)} =
    \left(
    W_h[:,i],\;
    W_{h+1}[i,:]
    \right).
\end{equation}
Given two networks, a cost matrix for hidden layer $h$ is constructed as
\begin{equation}
    C_{ij}^{(h)} = d\!\left(q_{i,A}^{(h)}, q_{j,B}^{(h)}\right),
\end{equation}
where $d$ is one of the structural distance metrics defined below.  Solving this assignment problem with the Hungarian algorithm yields a permutation of hidden layer $h$ in the second network.  This permutation is then applied simultaneously to the columns of $W_h$ and rows of $W_{h+1}$.

Because changing one hidden layer changes the profiles of adjacent hidden layers, a single pass is generally insufficient.  We therefore iterate over hidden layers in a zig-zag fashion.  For each hidden layer index $h$, the algorithm first sweeps backward through layers $h,h-1,\ldots,0$, and then forward through layers $0,1,\ldots,h$.  This sequence is repeated for several full passes.  After each pass, the full matched network distance is evaluated.  The procedure stops when the hidden-layer orders no longer change, when the improvement in the best distance falls below a tolerance, or when a maximum number of passes is reached.

The final distance between two aligned networks is a size-weighted average over layerwise matrix distances,
\begin{equation}
    D(\mathcal{W}_A,\mathcal{W}_B)
    =
    \sum_{\ell=0}^{L-1}
    \frac{|W_\ell|}{\sum_m |W_m|}
    d(W_{\ell,A},W_{\ell,B}),
\end{equation}
where $|W_\ell|$ denotes the number of entries in matrix $W_\ell$.  Importantly, the same local metric $d$ is used both to construct the Hungarian cost matrices and to compute the final matched network distance.

\subsection{Structural network distance metrics}

All local metrics compare two equally shaped matrices or vectors $A$ and $B$.  In this work we use three metrics: Euclidean distance, SND, and Alpha distance.

\paragraph{Euclidean distance.}
The Euclidean baseline is the root-mean-square entrywise difference,
\begin{equation}
    d_{\mathrm{Eucl}}(A,B)
    =
    \sqrt{
    \frac{1}{N}
    \sum_{ij}
    \left(A_{ij}-B_{ij}\right)^2
    },
\end{equation}
where $N$ is the number of matrix entries.

\paragraph{Structural Network Dissimilarity.}
The SND metric compares signed weights in a relative manner.  For each pair of entries $a=A_{ij}$ and $b=B_{ij}$, the local contribution is
\begin{equation}
    \delta_{\mathrm{SND}}(a,b)
    =
    \begin{cases}
    0, & a=0 \;\mathrm{and}\; b=0,\\
    1, & \operatorname{sgn}(a) \neq \operatorname{sgn}(b),\\
    \left|\dfrac{a-b}{a+b}\right|, & \operatorname{sgn}(a)=\operatorname{sgn}(b).
    \end{cases}
\end{equation}
The matrix distance is the mean over all entries,
\begin{equation}
    d_{\mathrm{SND}}(A,B)
    =
    \frac{1}{N}
    \sum_{ij}
    \delta_{\mathrm{SND}}(A_{ij},B_{ij}).
\end{equation}
Thus, sign mismatches are penalized strongly, while same-sign weights are compared by their relative difference.

\paragraph{Alpha distance.}
The Alpha distance is an importance-weighted sign mismatch measure.  For each entry pair $a=A_{ij}$ and $b=B_{ij}$, we define
\begin{equation}
    p_{ij}
    =
    \frac{
    |A_{ij}B_{ij}|^{\alpha}
    }{
    \sum_{mn} |A_{mn}B_{mn}|^{\alpha}
    },
\end{equation}
and
\begin{equation}
    S_{ij}
    =
    \begin{cases}
    1, & \operatorname{sgn}(A_{ij}) \neq \operatorname{sgn}(B_{ij}),\\
    0, & \mathrm{otherwise}.
    \end{cases}
\end{equation}
The Alpha distance is then
\begin{equation}
    d_{\alpha}(A,B)
    =
    \sum_{ij} p_{ij} S_{ij}.
\end{equation}
This metric ignores sign mismatches of entries that are small in at least one of the two matrices, and emphasizes sign mismatches of jointly large entries.  In the experiments below, we use $\alpha=2$ unless stated otherwise.

\subsection{Network matching test}

Before applying the matching procedure to trained networks, we tested it on a synthetic control problem in which the correct alignment was known by construction.  We generated an untrained multilayer perceptron with architecture
\begin{equation}
    20 \to 12 \to 10 \to 8 \to 6 .
\end{equation}
Its four weight matrices were sampled independently from zero-mean Gaussian distributions, with layer-dependent standard deviation
\begin{equation}
    \sigma_\ell = \sqrt{\frac{2}{n_{\ell}+n_{\ell+1}}},
\end{equation}
where $n_\ell$ and $n_{\ell+1}$ are the input and output sizes of layer $\ell$.

From this original network, denoted Net-A, we constructed a perturbed network Net-B in two steps.  First, each hidden layer was randomly permuted.  Importantly, these were true neuron permutations: permuting a hidden layer means permuting the corresponding columns of the incoming weight matrix and the corresponding rows of the outgoing weight matrix.  Second, independent Gaussian noise was added to every weight matrix, with noise standard deviation equal to 30\% of the standard deviation of the corresponding original matrix.

\begin{figure}[p]
\centering
\includegraphics[width=\linewidth]{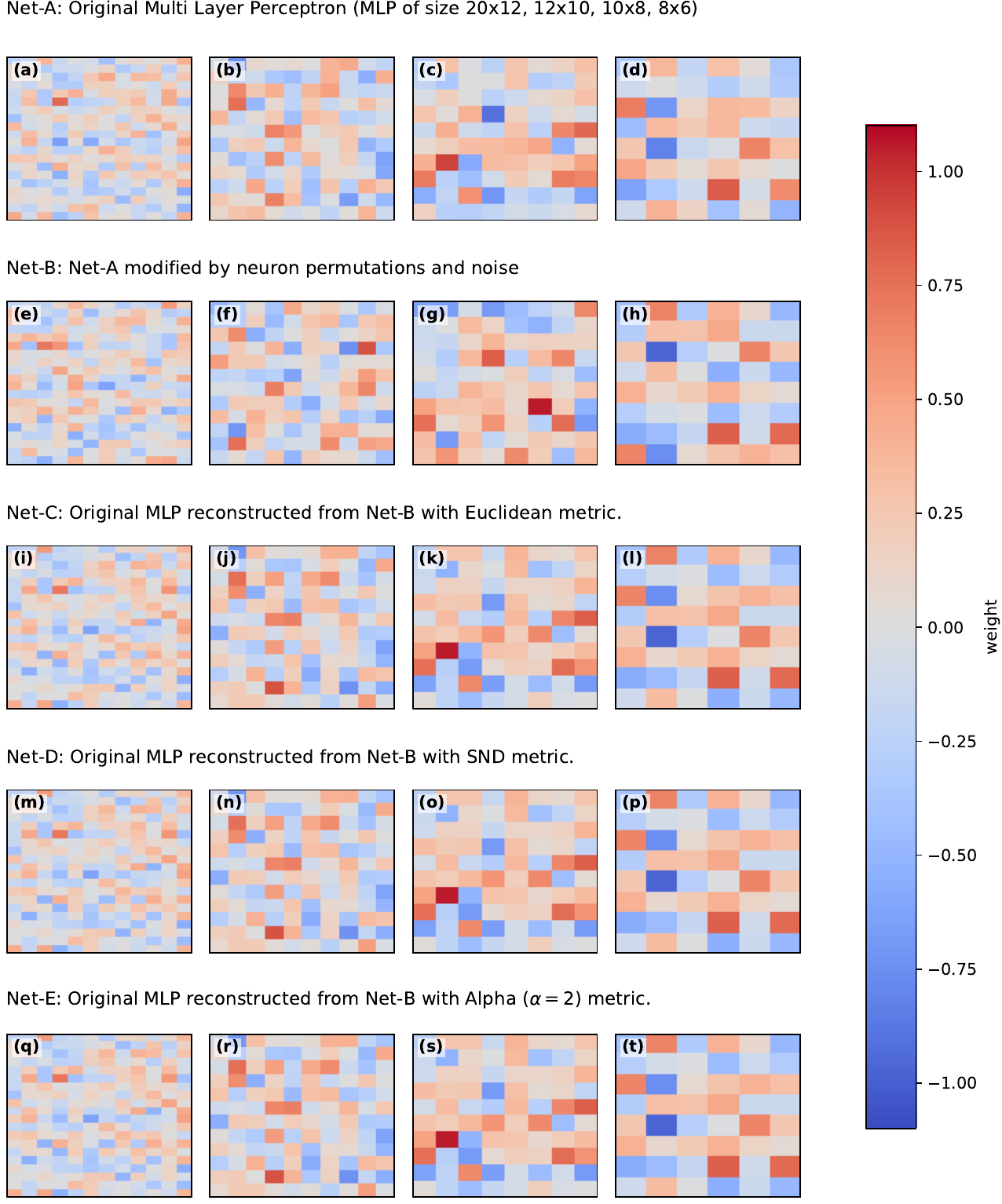}
\caption{
{\bf Matching demonstration for a synthetic multilayer perceptron.}
A randomly initialized multilayer perceptron of size $20 \to 12 \to 10 \to 8 \to 6$ was modified by hidden-neuron permutations and additive noise.
The first two rows show the original network and the perturbed network.
The following rows show the network reconstructed from the perturbed version by matching using Euclidean, SND, and Alpha distance with $\alpha=2$.
}
\label{FIG_MATCH_DEMO_VALS}
\end{figure}

\begin{figure}[p]
\centering
\includegraphics[width=\linewidth]{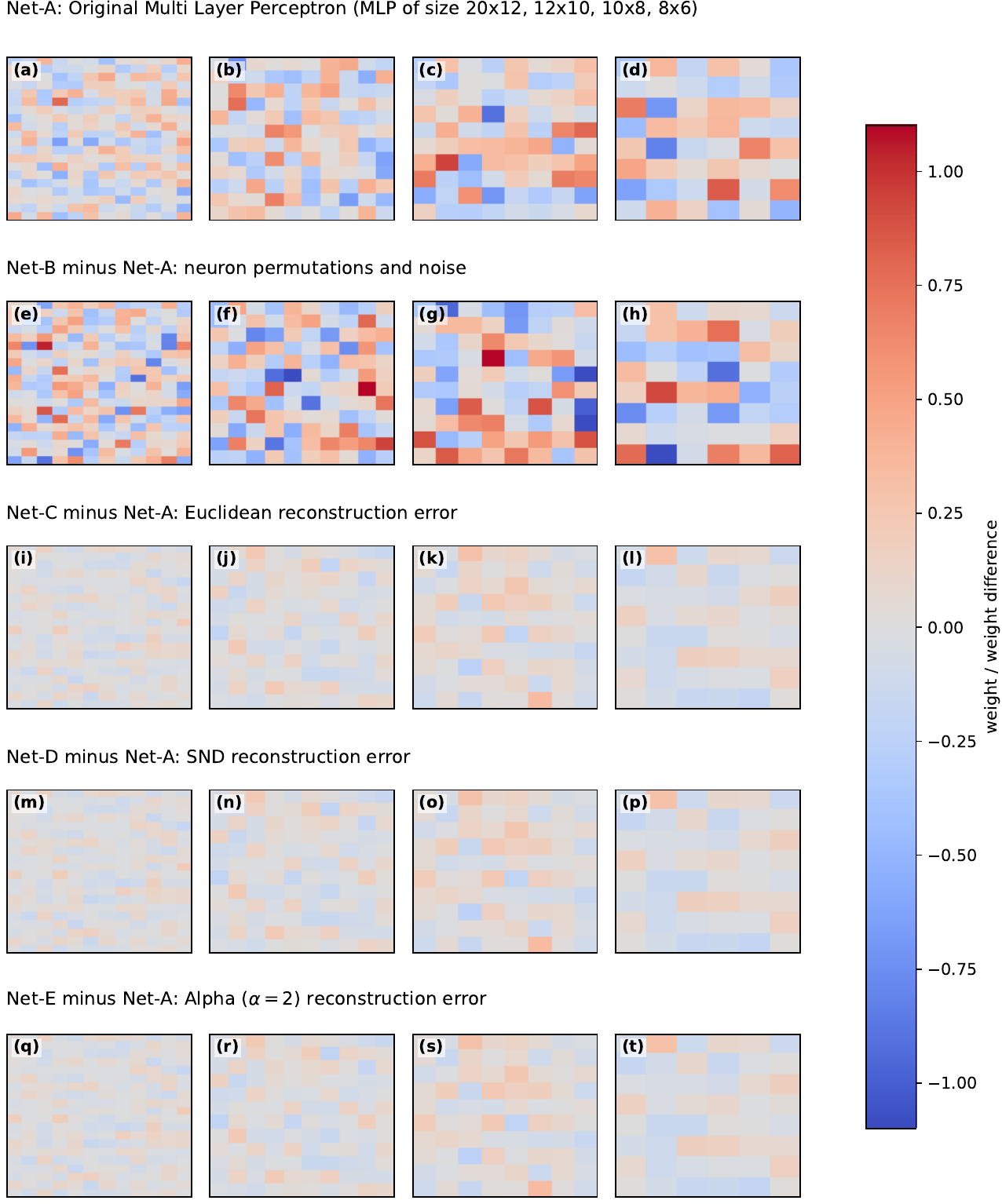}
\caption{
{\bf Reconstruction errors for the synthetic matching demonstration.}
The same experiment as in Fig.~\ref{FIG_MATCH_DEMO_VALS} is shown, but the perturbed and reconstructed networks are displayed as differences relative to the original network.
This representation highlights which matrix entries remain mismatched after reconstruction with Euclidean, SND, and Alpha distance with $\alpha=2$.
}
\label{FIG_MATCH_DEMO_DIFF}
\end{figure}

We then matched Net-B back to Net-A using the full network matching procedure described above, separately for the Euclidean, SND, and Alpha metrics.  In all three cases, PIF-based presorting was enabled, iterative Hungarian matching was allowed for up to 30 passes, and Alpha was evaluated with $\alpha=2$.  Figure~\ref{FIG_MATCH_DEMO_VALS} shows the original network, the perturbed network, and the three metric-dependent reconstructions.  Figure~\ref{FIG_MATCH_DEMO_DIFF} shows the same result as matrix differences relative to Net-A.  This test is not intended as a training experiment; it is a technical validation that the matching procedure can recover a known hidden-neuron alignment in a multilayer setting, up to the deliberately added noise.

\subsection{Data visualization with multidimensional scaling}

To visualize the geometry of trained networks, we used metric multidimensional scaling (MDS).  For a fixed training time point, we first computed the full pairwise distance matrix between the nine networks using the matched network distance described above.  MDS was then applied to this precomputed distance matrix to obtain a two-dimensional point configuration.  The goal of metric MDS is to find planar coordinates $x_i \in \mathbb{R}^2$ whose Euclidean distances approximate the original network distances $D_{ij}$.  In qualitative terms, MDS minimizes a stress function of the form
\begin{equation}
    \mathrm{stress}
    =
    \sum_{i<j}
    \left(
    \|x_i-x_j\|_2 - D_{ij}
    \right)^2 .
\end{equation}

All MDS plots in this study were computed independently for the selected training time points, using the corresponding distance matrix at that time point.  We used metric MDS with two output dimensions and precomputed dissimilarities.  To make the resulting plots visually comparable across panels, each two-dimensional configuration was centered and then rotated or reflected into a common orientation using a singular-value decomposition of the task-center configuration.  This postprocessing changes only the coordinate frame of the visualization; it does not change the relative distances among points within a panel.

The MDS axes have no intrinsic physical meaning and were therefore not labeled numerically.  The purpose of these plots is to provide an intuitive low-dimensional view of the task-dependent network geometry, in particular the separation between networks trained on MNIST, Fashion-MNIST, and KMNIST.

\section{Results}

\subsection{Demonstration of network matching}

\NI Since our subsequent experiments depend critically on resolving the neuron permutation problem, we first validate our network-matching method using a synthetic test case (for details, see the Methods section).

\NI To this end, we consider a multilayer perceptron (MLP) with layer sizes $20\rightarrow12\rightarrow10\rightarrow8\rightarrow6$. The four interlayer weight matrices are initialized independently with random values (Fig.\ref{FIG_MATCH_DEMO_VALS}, panels a–d). We then randomly permute the neurons in the two hidden layers and add noise to all matrix elements (panels e–h). As a result, the two scrambled hidden-layer representations are no longer directly recognizable as being structurally very similar to the original matrices.

\NI We then apply our network-matching method to reconstruct the original matrices from the scrambled ones, comparing three different metrics: Euclidean distance (panels i–l), SND (panels m–p), and Alpha (panels q–t).

\NI For all three metrics, the original weight matrices are recovered from the scrambled matrices with a high degree of accuracy. This becomes even more apparent when considering the reconstruction errors rather than the raw matrix values (Fig.\ref{FIG_MATCH_DEMO_DIFF}, rows three to five).

\NI The network-matching method has also been tested and evaluated systematically in previous work (CITE THESIS; data not shown).

\subsection{Network distances during training}

\NI Next, we consider smaller MLPs with a single hidden layer and sizes $784\rightarrow64\rightarrow10$. In this architecture, each of the two weight matrices is directly connected to either the fixed input layer or the fixed output layer, making the matching problem considerably simpler. Nevertheless, matching remains essential for comparing the structural similarity of independently trained networks whose hidden-layer neurons are arbitrarily permuted.

\NI For this experiment, we use nine such MLPs, each initialized independently with different random weights. The networks are divided into three groups, and each group is trained on a different classification task: MNIST (M), Fashion-MNIST (F), or KMNIST (K). These datasets share the same input dimensionality, the same number of classes, and the same numbers of training and test samples.

\NI After each training epoch, we match all nine networks pairwise using our method and compute the corresponding pairwise network distances. These distances are then aggregated into six representative quantities: the within-task distances MM, FF, and KK, and the between-task distances MF, MK, and FK. We then examine how these six aggregated distances evolve over the course of training.

\begin{figure}[p]
\centering
\includegraphics[width=\linewidth]{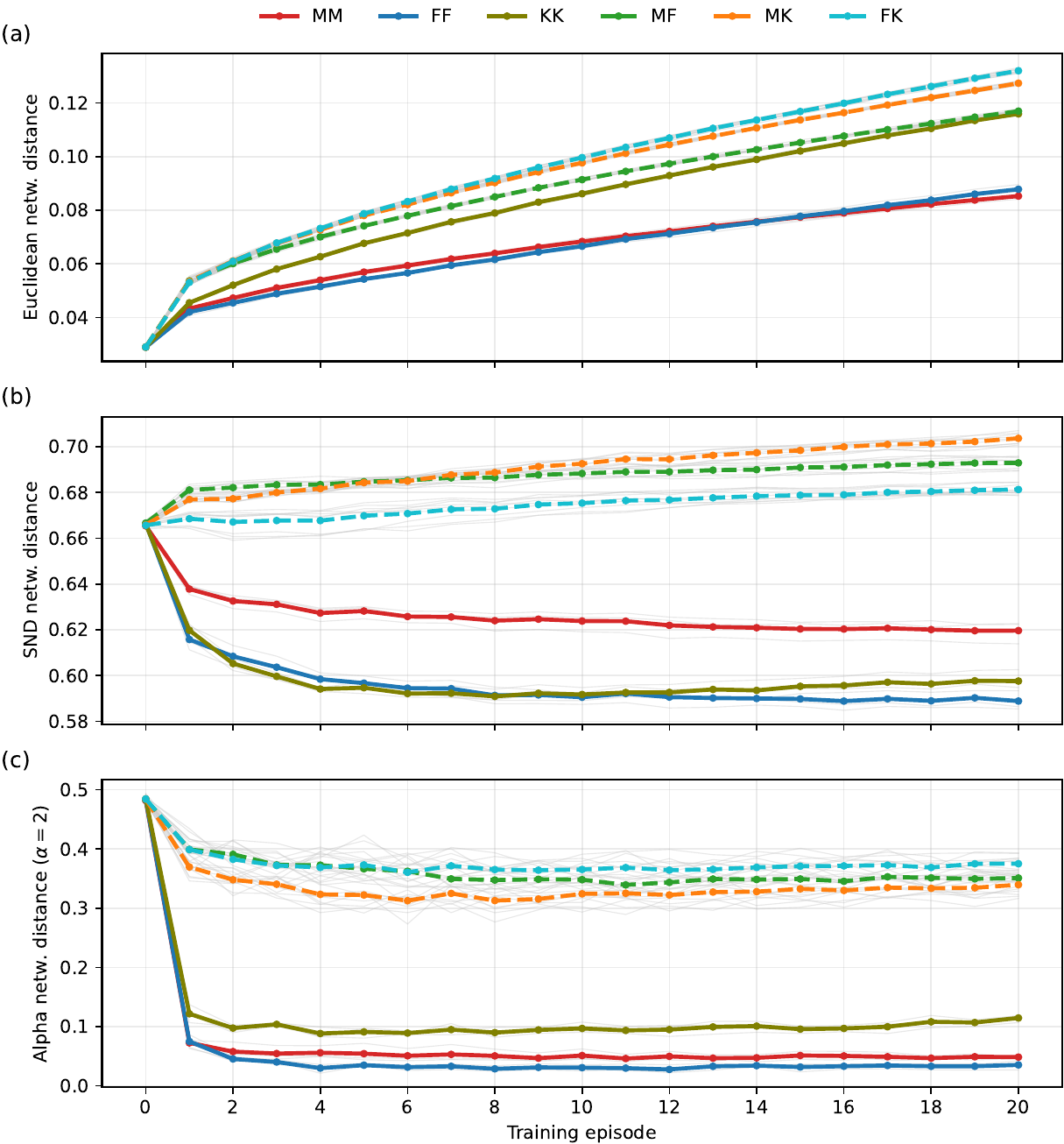}
\caption{
{\bf Network distances during training on MNIST, Fashion, and KMNIST.}
Nine small MLPs of size $784 \to 64 \to 10$ were trained, three per task, with independent random seeds.
Pairwise matched network distances were computed after each training episode using Euclidean distance, SND, and Alpha distance with $\alpha=2$.
Light gray curves show individual pair distances, while colored curves show group means for within-task and between-task comparisons.
MNIST, Fashion, and KMNIST are shown in red, blue, and olive, respectively.
}
\label{FIG_DIST_VERSUS_EPOCH}
\end{figure}

\NI This analysis is performed independently for each of our three metrics: Euclidean distance (Fig.\ref{FIG_DIST_VERSUS_EPOCH}, panel (a)), SND (panel (b)), and Alpha (panel (c)).

\NI For all three metrics, the six distance curves begin at approximately the same value at epoch 0, before training has started. At first sight, this may seem surprising, since all networks were initialized independently with uniformly distributed random weights. However, the network distances ultimately represent averages of local differences between corresponding matrix elements. Because the matrices contain a very large number of elements, these averages exhibit only small fluctuations and are therefore highly stable across independently initialized networks.

\NI With the Euclidean metric, all six representative distances increase as training progresses. This largely reflects the metric’s sensitivity to global changes in the weight matrices, such as an increase in the standard deviation of the weight distributions. Moreover, the Euclidean metric does not strongly distinguish between matrix elements of small magnitude, which are likely to be of limited relevance to the task, and dominant elements of large magnitude. Nevertheless, even under the Euclidean metric, the within-task distances generally remain smaller than the between-task distances, although the KK curve eventually approaches and touches the MF curve near the end of training.

\NI Under the SND metric, by contrast, the between-task distances increase over the course of training, while the within-task distances decrease. This demonstrates that SND is better suited to capturing the emergence of task-specific structural similarities among networks trained on the same task. The metric is relatively insensitive to differences in the overall scale of the weight distributions and strongly penalizes corresponding matrix elements with opposite signs.
However, the total change in the SND distances over the entire training process remains comparatively small. Consequently, visualization methods that represent each network as a point in a two-dimensional plane are unlikely to reveal a pronounced clustering structure when applied to SND-based distance matrices.

\NI With the Alpha metric, using $\alpha=2$ throughout this paper, all six distances decrease rapidly during the first training epoch and then continue to decline much more gradually. This metric is invariant under independent rescaling of the two networks or matrices being compared and places particular emphasis on matrix elements of large magnitude. It produces both large overall changes in distance and a clear separation between the within-task and between-task curves. The Alpha metric is therefore particularly well suited to revealing a possible convergence of learning trajectories toward task-specific regions in the space of network structures.

\begin{figure}[p]
\centering
\includegraphics[width=0.78\linewidth]{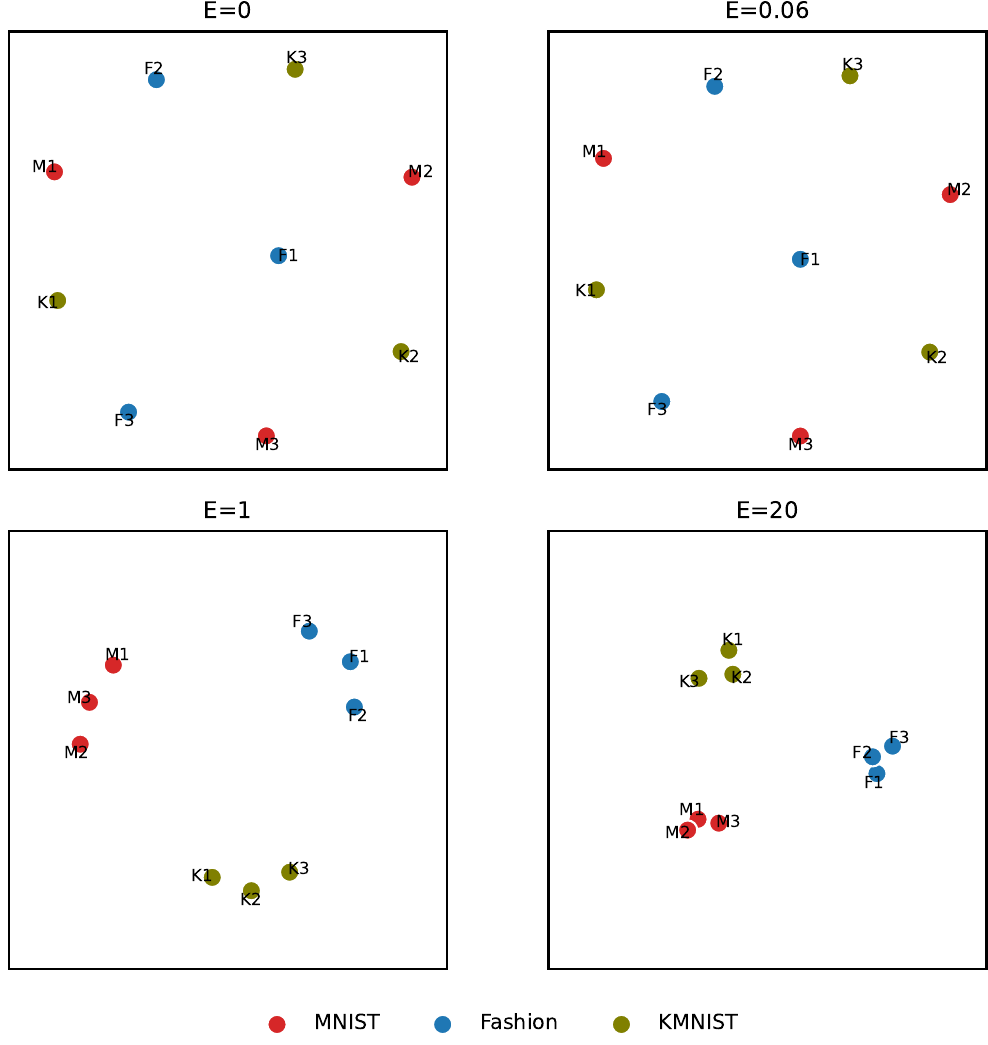}
\caption{
{\bf Early MDS visualization of Alpha network distances.}
Metric multidimensional scaling was applied to Alpha-distance matrices with $\alpha=2$ for the same nine trained MLPs as in Fig.~\ref{FIG_DIST_VERSUS_EPOCH}.
The panels show the network configuration before training ($E=0$), at the onset of measurable between-task separation ($E=0.06$), after approximately one training episode ($E=1$), and after the final episode ($E=20$).
Here $E=0.06$ is the same checkpoint marked by the vertical dashed line in Fig.~\ref{FIG_EARLY_EVO}; in the stored training trajectory it corresponds to mini-batch update 7, i.e. $E \approx 0.0596$.
The task-specific clusters are not yet visually separated at this early onset point, but are already clearly separated by about one training episode.
}
\label{FIG_MDS}
\end{figure}

\NI To examine this convergence more directly, we visualize the distribution of networks in structural space using multidimensional scaling (MDS). At four selected stages of training, we compute the $9\times9$ matrix of pairwise Alpha distances between all networks. For each snapshot, MDS then places the nine networks as points in a two-dimensional plane while approximately preserving their mutual distances (Fig.\ref{FIG_MDS}). Before training and shortly after its onset ($E=0$ and $E=0.06$), the nine networks form a broad, unstructured configuration. After the first complete training epoch ($E=1$), however, three distinct clusters have already emerged, corresponding to the three training tasks. During the remainder of training, these clusters become progressively more compact.

\subsection{High temporal resolution of network distances and accuracy}

\begin{figure}[p]
\centering
\includegraphics[width=\linewidth]{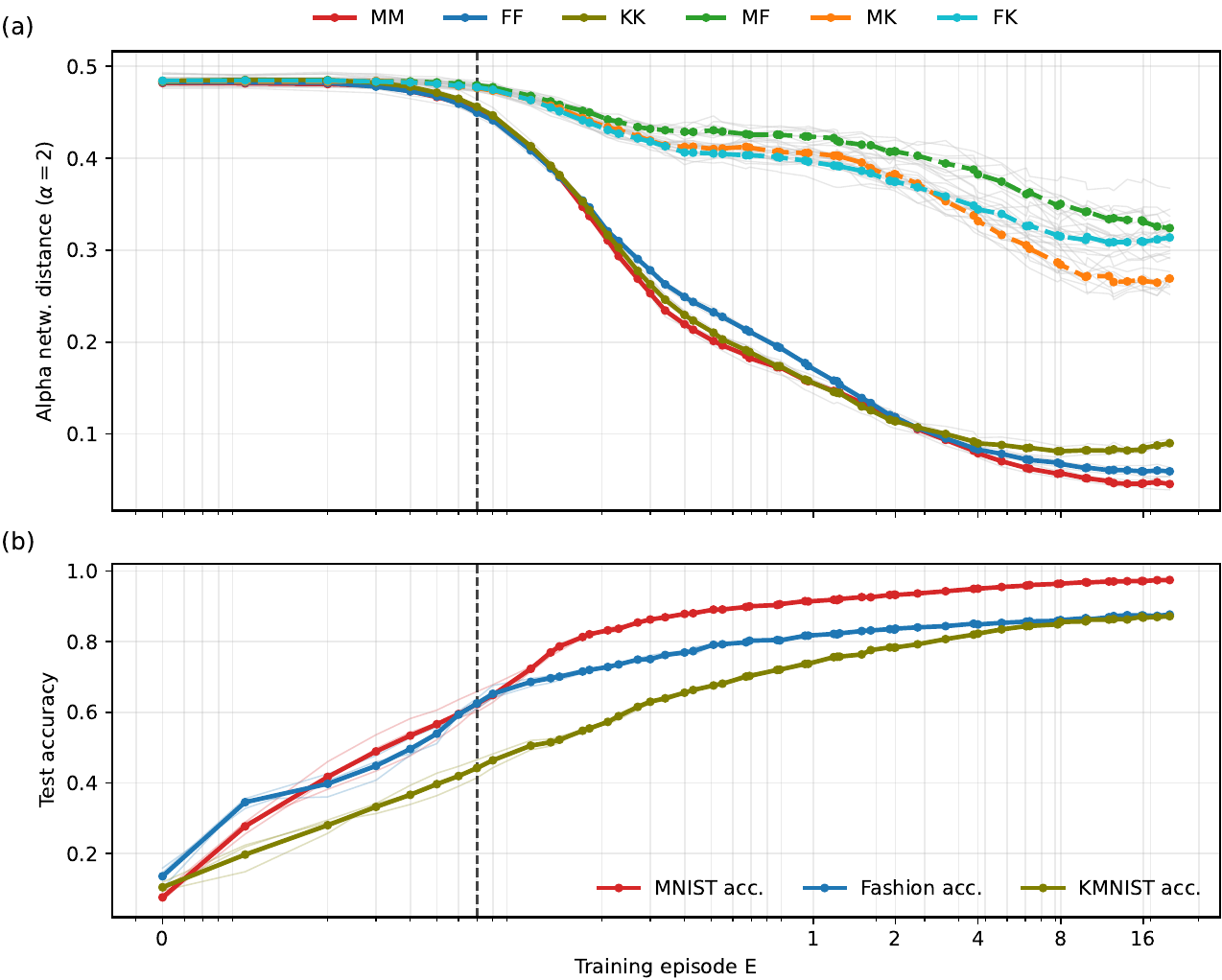}
\caption{
{\bf Early training dynamics of Alpha network distances and task performance.}
The same three-task setup as in Fig.~\ref{FIG_DIST_VERSUS_EPOCH} was repeated with logarithmically spaced checkpoints during training.
The upper panel shows matched Alpha network distances with $\alpha=2$, while the lower panel shows the corresponding test accuracies.
The horizontal axis is logarithmic in optimizer updates, but tick labels indicate the corresponding training episode $E$.
The vertical dashed line marks the first saved checkpoint at which the mean between-task distance exceeds the mean within-task distance by more than 0.02 ($E \approx 0.06$).
This representation resolves the rapid initial changes that are compressed in the uniformly sampled episode plots.
}
\label{FIG_EARLY_EVO}
\end{figure}

\NI The preceding results revealed a pronounced decrease in the within-task Alpha distances, particularly during the very first training epoch. We therefore recompute the evolution of the network distances at a higher temporal resolution (Fig.\ref{FIG_EARLY_EVO}, panel (a)) and additionally examine the test accuracies as functions of training time (panel (b)).

\NI We find that the mutual structural distances among the nine networks remain nearly identical and approximately constant until about $E=0.06$, when a gap begins to emerge between the mean between-task and within-task distances. During this early phase, all three tasks exhibit very similar behavior: the MM, FF, and KK curves nearly coincide, as do the MF, MK, and FK curves. Beyond $E=0.06$, the separation between the between-task and within-task distances increases substantially, while task-specific differences also begin to appear.

\NI As expected, the accuracies achieved on all three tasks increase monotonically over the course of training. By the end of epoch 20, the simpler MNIST task reaches a higher performance level than Fashion-MNIST or KMNIST. Remarkably, however, the accuracies already rise far above chance before the critical time point $E=0.06$, at which the structural distances begin to separate and decrease; for MNIST, the accuracy has already reached 0.623 at this point. This suggests that, during the earliest stage of training, the weights undergo subtle but functionally important changes that strongly improve performance while remaining largely undetected by the Alpha distance metric.

\subsection{Importance of large-magnitude weights}

\NI By using the Alpha metric with an exponent of $\alpha=2$, which nonlinearly increases the relative contribution of weights with large magnitudes, we implicitly assume that these strong weights are particularly relevant to the learned task. To test this assumption, we perform ablation experiments on three MLPs trained on MNIST.

\NI We first set progressively larger fractions of the weakest weights to zero and examine the resulting decrease in test accuracy. We then perform the complementary experiment by progressively removing the strongest weights. Both analyses are carried out separately for the first weight matrix, $W_0$, and the final weight matrix, $W_1$, of each MLP.

\begin{figure}[p]
\centering
\includegraphics[width=\linewidth]{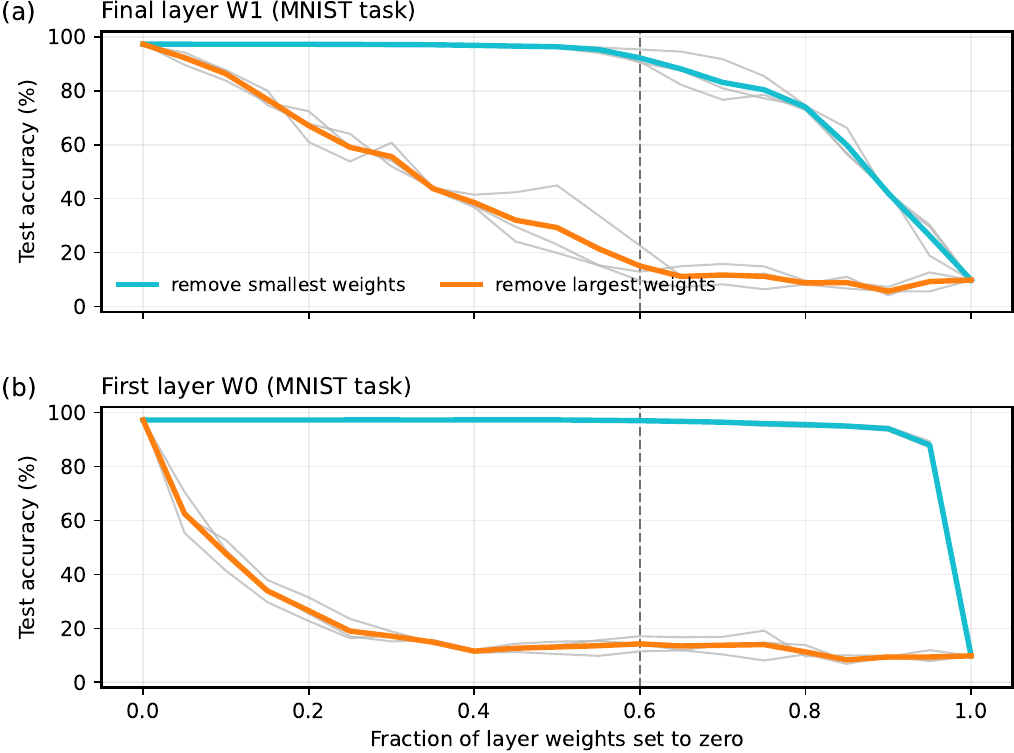}
\caption{
{\bf Functional importance of large weights in trained MNIST networks.}
For the three independently trained MNIST networks, weights were ablated in the final trained model by setting a specified fraction of weights in one layer to zero.
Panel (a) shows ablations in the hidden-to-output matrix $W_1$, and panel (b) shows the corresponding analysis for the input-to-hidden matrix $W_0$.
In each panel, the smallest weights by absolute value or the largest weights by absolute value were removed separately.
Gray curves show individual networks, while thick colored curves show the mean across the three runs.
Removing the largest weights rapidly destroys test performance, whereas removing the smallest weights has little effect until a large majority of weights has been removed.
The vertical dashed line marks removal of 60\% of the weights.
}
\label{FIG_WEIGHT_CUTOFF}
\end{figure}

\NI In the first layer, nearly 90\% of the weakest weights can be removed with little effect on test accuracy (Fig.\ref{FIG_WEIGHT_CUTOFF}, panel (b), blue curve). By contrast, removing only 10\% of the strongest weights already causes a dramatic collapse in performance (panel (b), orange curve).

\NI In the final layer, approximately 60\% of the weakest weights can be removed without substantially impairing performance (panel (a), blue curve). In this layer, however, removing the strongest weights produces a more gradual decline in accuracy (panel (a), orange curve).

\NI Overall, these results show that, in our test case, approximately 60\% of the weakest weights across the entire two-layer network can be removed without significantly affecting test accuracy. This finding supports the central intuition underlying the Alpha metric: weights of large magnitude tend to carry more task-relevant structural information.

\subsection{Task-relevance-dependent weight evolution}

\NI If the strongest weights are also the most important ones in the fully trained networks, this raises the question of how these weights emerge during training. Do the weights that ultimately become task-relevant pass through repeated phases of strengthening and weakening before settling into their final values?

\begin{figure}[p]
\centering
\includegraphics[width=\linewidth]{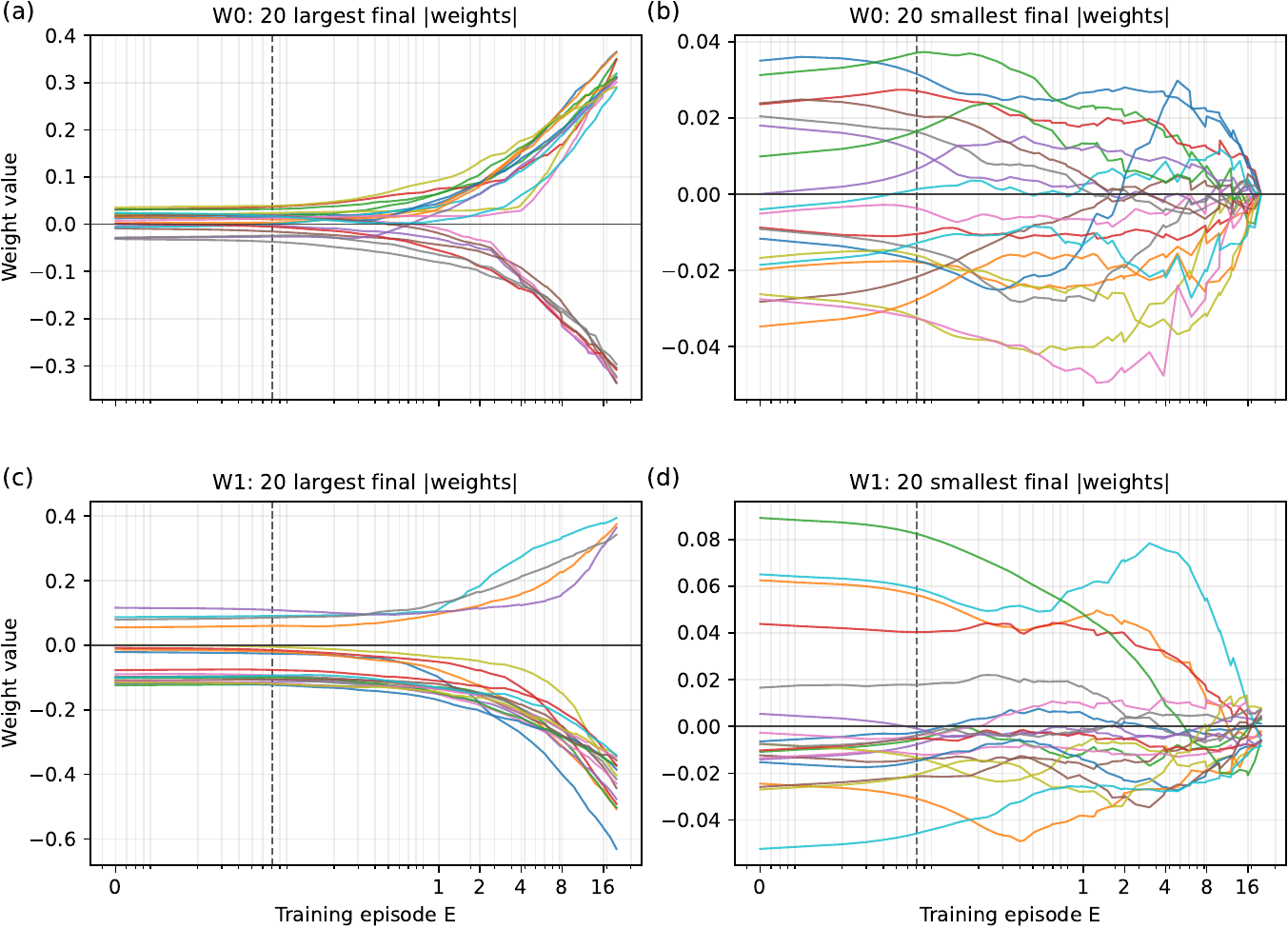}
\caption{
{\bf Evolution of selected individual weights during training.}
For one Fashion-trained MLP, the 20 weights with largest final absolute value and the 20 weights with smallest final absolute value were selected separately in both weight matrices.
The signed values of these fixed entries are shown over the same logarithmically sampled training checkpoints used in Fig.~\ref{FIG_EARLY_EVO}.
Panels (a,b) show the input-to-hidden matrix $W_0$, and panels (c,d) show the hidden-to-output matrix $W_1$.
The left panels track weights that become large by the end of training, whereas the right panels track weights that remain close to zero.
}
\label{FIG_WEIGHT_EVO}
\end{figure}

\NI To investigate this question, we identify the 20 strongest and the 20 weakest weights in a fully trained network and trace their values throughout the entire training period (Fig.\ref{FIG_WEIGHT_EVO}). In particular, we count how often each group of weights changes sign during training.

\NI In the first layer, $W_0$, we observe only 5 sign changes among the 20 strongest weights, compared with 40 sign changes among the 20 weakest weights. Similarly, in the final layer, $W_1$, we find only 1 sign change in the strong-weight group, but 25 sign changes in the weak-weight group.

\NI These results indicate that most of the weights that ultimately become highly task-relevant already possess their final sign in the randomly initialized network at epoch 0. During training, these pre-existing carriers of task-relevant structure are progressively amplified (Fig.\ref{FIG_WEIGHT_EVO}, panels (a) and (c)). By contrast, functionally less important weights follow a considerably less stable trajectory, frequently changing sign before eventually being suppressed toward small magnitudes (Fig.\ref{FIG_WEIGHT_EVO}, panels (b) and (d)).

\subsection{Two phases of structural development}

\NI We now return to our earlier observation that, during the initial phase of weight evolution, up to approximately $E=0.06$, test accuracy increases sharply even though the structural distances remain almost unchanged (Fig.\ref{FIG_EARLY_EVO}).

\NI In principle, this finding could reflect a limited sensitivity of the Alpha metric to very small changes in the weights. However, reconsidering the weight-tracing experiment shown in Fig.\ref{FIG_WEIGHT_EVO} reveals that the 20 most important weights indeed change very little during this early period (panels (a) and (c), before the dashed vertical line). By contrast, somewhat stronger dynamics are visible among the functionally less important weights (panels (b) and (d)).

\begin{figure}[p]
\centering
\includegraphics[width=\linewidth]{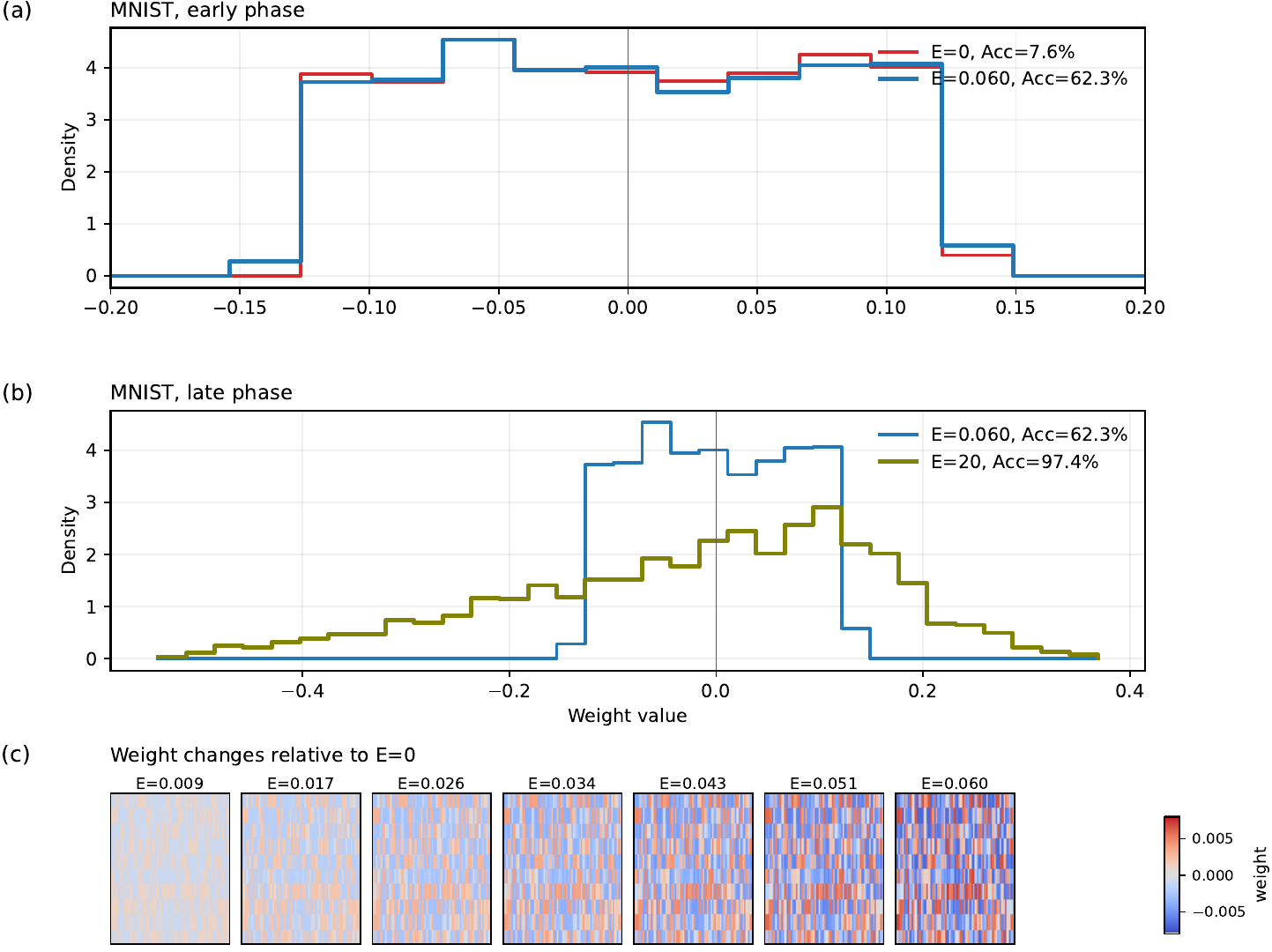}
\caption{
{\bf Early weight-distribution stability and elementwise changes.}
The hidden-to-output weight matrix $W_1$ of the MNIST-trained networks was analyzed during the early phase of training.
Panels (a,b) show pooled weight distributions across the three independently initialized MNIST runs.
Panel (a) compares the initial distribution ($E=0$) with the distribution at the early separation marker from Fig.~\ref{FIG_EARLY_EVO} ($E \approx 0.06$).
Panel (b) compares this same early distribution with the final distribution at $E=20$.
Although the distribution changes only weakly during the early phase, panel (c) shows that individual entries of $W_1$ already exhibit visible changes relative to $E=0$ in a representative run (M1).
Thus, early accuracy gains can occur while the global weight distribution remains nearly unchanged, even though individual matrix elements have begun to drift.
}
\label{FIG_WEIGHT_DISTRIBUTIONS}
\end{figure}

\NI To investigate this point further, we compute the probability density distributions of the weights in layer $W_1$ at initialization (Fig.\ref{FIG_WEIGHT_DISTRIBUTIONS}, panel (a), red curve), at the end of the early phase (panels (a) and (b), blue curves), and at the end of training (panel (b), olive curve). We find that the weight distribution changes only minimally during the early phase, whereas the late phase is characterized by a pronounced broadening and an asymmetric change in the shape of the distribution.

\NI The stability of the global weight distribution does not, of course, imply that the individual weights remain fixed. To illustrate this, we also plot the difference between the weight matrix $W_1(E)$ and its initial state, $W_1(E\!=\!0)$, at seven time points during the early phase (Fig.\ref{FIG_WEIGHT_DISTRIBUTIONS}, panel (c)). These plots reveal numerous small-amplitude changes distributed across the entire matrix. However, the changes are evidently arranged in such a way that the overall distribution of weight values remains nearly unchanged.

\subsection{Exploitation of random initial structures}

\NI Finally, we revisit the finding from Fig.\ref{FIG_WEIGHT_EVO} that most of the weights that ultimately become task-relevant already possess their final sign at initialization.

\NI This observation suggests not only that functionally relevant building blocks for the MNIST task are already present in the random initial weight matrix, but also that a suitably initialized random matrix may contain latent building blocks for a wide range of possible tasks. Training would then act primarily by selecting and amplifying those structures that are relevant to the task at hand.

\begin{figure}[p]
\centering
\includegraphics[width=0.86\linewidth]{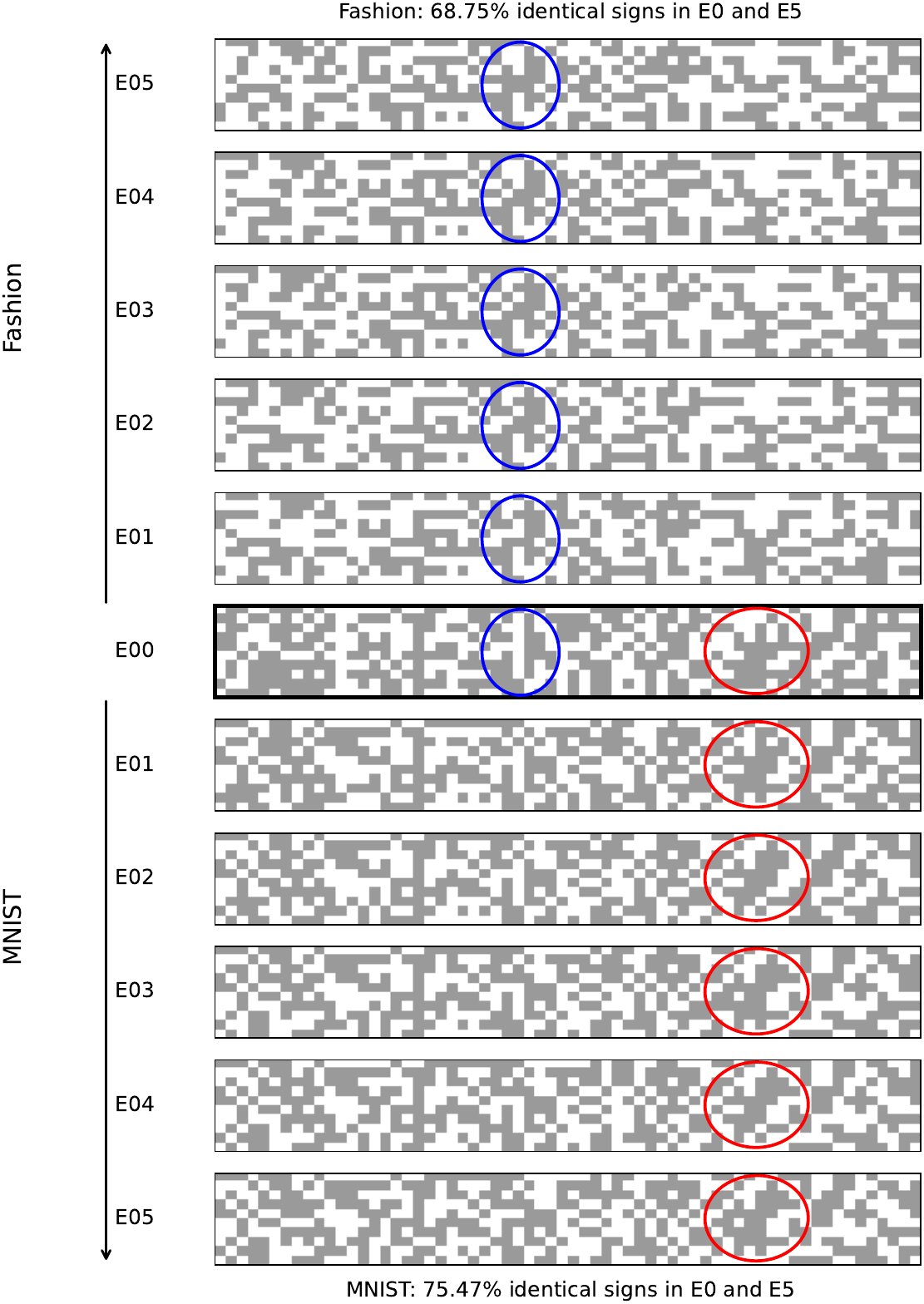}
\caption{
{\bf Divergent training trajectories from a common initial network.}
Two MLPs with identical initial weights were trained on Fashion and MNIST, respectively.
The figure shows the sign pattern of the hidden-to-output weight matrix $W_1$ over early training epochs, with the common initial matrix in the center and task-specific trajectories branching upward and downward.
Highlighted patterns illustrate matrix regions whose sign structure is already partially visible in the common initial network and is subsequently amplified or reshaped by task-specific training.
Between episode 0 and episode 5, 68.75\% of signs remained identical for Fashion and 75.47\% for MNIST.
}
\label{FIG_ONE_SEED_TWO_TASKS}
\end{figure}

\NI To test this hypothesis, we start from the same randomly initialized $W_1$ matrix and train the corresponding network once on MNIST and once on Fashion-MNIST. In both cases, the evolution of $W_1$ is shown explicitly for the first five training epochs (Fig.\ref{FIG_ONE_SEED_TWO_TASKS}). To facilitate visual pattern recognition, only the signs of the weights are displayed.

\NI In both training runs, several prominent local patterns of neighboring matrix elements can be identified by visual inspection. These patterns are already present at initialization, denoted by E00 in the center of the figure, and remain recognizable throughout the subsequent training process. However, some of the most functionally important structures may not be visually apparent in this representation: In layer $W_1$, the 64 matrix columns correspond to hidden-layer neurons whose ordering is arbitrary. Consequently, groups of functionally related weights may be distributed across widely separated columns rather than appearing as contiguous patterns within a matrix row.

\section{Discussion}

\subsection{Summary}

\NI In this study, we asked whether independently trained neural networks develop task-specific structural similarities that can be detected directly in weight space.  This question is complicated by the neuron permutation problem: two networks may implement closely related solutions while assigning corresponding functional roles to different hidden-neuron indices.  We therefore treated matching as an integral part of the network distance itself.  Our procedure first establishes a rough alignment by PIF-based presorting and then refines this alignment by iterative Hungarian matching, using the same structural metric for both local matching costs and final network-distance evaluation.

\NI Applying this framework to small multi-layer perceptrons trained on MNIST, Fashion-MNIST, and KMNIST revealed that structural convergence depends strongly on the metric used to compare matched networks.  Euclidean distance mainly captured the overall displacement of weights during training and therefore increased with training time.  SND and, more clearly, the Alpha distance emphasized structural agreement among important weights.  With Alpha distance, independently initialized networks trained on the same task moved toward more similar regions of weight space, whereas networks trained on different tasks remained more separated.  Thus, the apparent geometry of training is not a metric-independent property; it depends on which aspects of the weight matrices are regarded as structurally meaningful.

\NI Several control analyses support the interpretation that large-magnitude weights carry disproportionate task-relevant information.  Direct ablation showed that removing the largest weights from either layer rapidly destroyed MNIST performance, whereas removing even a large fraction of the smallest weights had little effect.  Likewise, tracking selected individual weights showed that weights that became large by the end of training were relatively sign-stable, while weights that remained small fluctuated much more often around zero.  These observations provide an empirical rationale for the Alpha metric, which discounts weak weights and focuses on sign agreement among weights that are large in both compared networks.

\NI A particularly striking result was the existence of a very early functional learning phase.  Test accuracy rose strongly within the first small fraction of a training epoch, and the cross-entropy loss decreased smoothly and substantially over the same interval.  At the same time, the weight matrices changed only very little: the global distribution of weights remained almost unchanged, and even the directly visible matrix differences were on the order of only a few thousandths.  A simple control in which comparable uniform random perturbations were added to the initial matrix confirmed that such small changes are too small to visibly alter the global weight distribution.  Thus, early learning is not accompanied by a large morphological reorganization of the weight matrices.  Instead, small but coherent parameter updates already produce large functional consequences.

\NI This suggests a two-stage view of training.  In the earliest phase, the network undergoes a highly effective functional calibration: tiny, directed changes in weights move many samples toward or across correct decision boundaries.  Only later does a clearer structural reorganization become visible in the matched network distances, weight trajectories, and task-specific clustering.  In this sense, the initial random network is not an inert blank slate.  It already contains many random feature directions and partial structures that can be rapidly exploited by gradient descent before they are more slowly consolidated into task-specific weight patterns.

\NI Finally, the results are consistent with the idea that neural-network loss landscapes may be more training-friendly than a naive picture of isolated optima would suggest.  Because hidden neurons are freely permutable, any functional solution has a very large number of symmetry-equivalent realizations in parameter space.  Additional continuous redundancies and flat directions further enlarge these solution sets.  The landscape may therefore be effectively ``target-rich'': a random initialization need not be close to one unique canonical solution, but may already lie within the basin of attraction of one of many equivalent or nearly equivalent realizations.  This could help explain why the earliest gradients are already coherent and functionally effective, even though the associated movements in raw weight space remain extremely small.

\subsection{Open Questions and Outlook}

\paragraph{Scaling to richer visual tasks.}
The present study deliberately used small multi-layer perceptrons and relatively simple image-classification tasks.  This made it possible to compute matched network distances many times, to inspect individual weight matrices directly, and to test methodological assumptions under controlled conditions.  A first open question is therefore how far the observed signatures of task-specific structural convergence persist in larger networks and more complex datasets.  Natural next steps include convolutional architectures, deeper fully connected networks, and image datasets with greater visual variability.  Such experiments would test whether the Alpha-distance separation observed here is a peculiarity of small MLPs or whether it captures a more general form of task-induced structural convergence.

\paragraph{Datasets with explicit hierarchical structure.}
A second direction is to move beyond standard benchmark labels and construct datasets in which the latent task structure is known by design.  For example, one could generate synthetic image families with independently controlled factors such as shape, position, texture, contrast, compositional hierarchy, and class membership.  This would make it possible to ask whether networks trained on related subtasks converge first toward shared low-level structures and only later separate into task-specific higher-level structures.  Such controlled datasets could also reveal whether network trajectories contain recurring developmental phases, analogous in spirit to staged morphogenetic processes, rather than merely smooth movement toward a final classifier.

\paragraph{The early learning phase.}
The most surprising dynamical observation in the present work was the very early rise in accuracy and fall in loss despite extremely small changes in the weights.  This phase deserves a dedicated follow-up study.  Important questions include whether early improvements are driven mainly by changes in output-layer calibration, hidden-unit activation scales, coarse contrast features, or small coherent rotations of many decision boundaries.  The present results suggest that global weight distributions and even the largest final weights are too coarse to explain this phase by themselves.  Future analyses should therefore combine parameter-space measurements with activation statistics, gradient directions, class-margin dynamics, and sample-wise changes in prediction confidence.

\paragraph{Attractor geometry and basins of convergence.}
The idea of an algorithmic attractor remains only partially characterized by the present experiments.  We measured distances between independently trained endpoint networks and short training trajectories, but did not yet map the surrounding basins of attraction.  Future work could initialize networks along controlled interpolations between seeds, perturb trained networks in selected directions, or restart training from matched intermediate states.  Such experiments would help distinguish several possibilities: compact task-specific basins, extended low-loss manifolds, multiple discrete solution families within one task, or hierarchical attractor structures in which broad task families share early features but diverge later.

\paragraph{Language-processing attractors.}
The attractor idea is not limited to image-classification networks.  A particularly interesting extension is language processing, where one can ask whether meanings, propositions, or semantic roles form reproducible attractor-like regions in representation space.  Preliminary explorations in our project used the sentence-embedding model \texttt{all-mpnet-base-v2}, a transformer-based model related to BERT-style contextual representations and sentence-transformer training \cite{devlin2019bert,reimers2019sentencebert,song2020mpnet}.  In a small manually constructed pilot set, paraphrases of the same event were closer in cosine similarity than sentences from different event categories.  A simple evolutionary search over short word sequences could also increase similarity to a target embedding, although the resulting strings were not necessarily grammatical or semantically interpretable.  These observations are best regarded as motivation rather than evidence.  They suggest a future research program in which sentence embeddings, generated sentences, human meaning judgments, and fluency constraints are combined to ask whether semantic targets can be approached as attractors in a linguistic representation space.

\paragraph{Toward cross-level convergence.}
Ultimately, the broader goal is to connect convergence across several levels of description.  At the lowest level, one can compare matched weight matrices as done here.  At an intermediate level, one can compare activations, representational geometries, and decision boundaries.  At a higher level, one can ask whether different learning systems converge toward similar semantic or algorithmic descriptions of the same problem.  The present work provides one small but concrete step in this direction: it shows that, once permutation symmetries are handled, task-dependent convergence can be detected directly in parameter space.  The open challenge is to determine how such parameter-space signatures relate to representation-space similarity, functional robustness, and interpretable algorithmic structure.

\subsection{Relation to other Works}

\paragraph{Convergent evolution and algorithmic attractors.}
The conceptual starting point of this work is the analogy between biological convergent evolution and convergence in learning systems. In biology, convergent evolution is usually interpreted as evidence that functional constraints can make certain solutions recurrently accessible despite different histories and implementations \cite{conwayMorris2003lifes,losos2017improbable,petryshen2020evidence}. Our study transfers this idea to neural-network training. Independent training runs correspond to repeated developmental or evolutionary trials, while the task defines the functional constraint. The central question is whether the same task repeatedly guides networks toward similar internal organizations. In this sense, our proposed notion of an algorithmic attractor is closer to a dynamical and structural hypothesis than to a purely metaphorical analogy: it asks whether task constraints carve out preferred regions in the space of learnable algorithms.

Classical biological examples illustrate that convergence must be distinguished across levels of description. Sharks and dolphins converge strongly in overall body shape and hydrodynamic function, although their evolutionary histories, developmental pathways, and detailed anatomical implementations differ substantially. Birds and bats likewise converged independently on the functional solution of powered flight, but their wings are constructed differently: bird wings consist of modified forelimbs bearing feathers, whereas bat wings are formed by skin membranes stretched across highly elongated digits. Convergence therefore does not require identity of implementation. It may occur at the level of function, organization, or an abstract solution principle while leaving lower-level mechanisms substantially different.

This distinction maps naturally onto Marr's levels of analysis. Systems implemented in fundamentally different physical substrates may face the same functional or computational problem and nevertheless converge toward similar, or even effectively identical, algorithmic solutions. Sharks and dolphins, for example, realize efficient locomotion through water through very different biological substrates and developmental histories, yet converge on a closely related hydrodynamic organization. Birds and bats likewise implement powered flight through anatomically distinct wing structures, while sharing important aerodynamic solution principles. Convergence can therefore occur at the algorithmic level even when the implementational level remains substantially different.

Applied to neural networks, this implies that several forms of convergence must be distinguished. Functional convergence means that independently trained systems achieve similar input--output behavior. Representational convergence means that they develop similar activation geometries or latent variables. Algorithmic convergence means that they rely on similar internal transformations, organizational principles, or solution strategies. Parameter-space convergence, finally, means that, after accounting for symmetries such as hidden-neuron permutations, their learned weights exhibit reproducible structural similarities. Our notion of an algorithmic attractor expresses the hypothesis that a shared computational problem constrains the space of possible solutions strongly enough that independently developing systems converge toward related algorithmic organizations, even when their physical substrates, exact numerical parameters, and developmental or learning trajectories differ.

\paragraph{Representation similarity and convergent learning.}
The most direct technical neighbors of our work are studies that compare representations learned by independently trained networks.  Li et al. introduced the term convergent learning and showed that separately trained networks can learn features or feature subspaces that partially align across runs \cite{li2015convergentLearning}.  Subsequent work developed increasingly sophisticated methods for comparing activation spaces, including SVCCA \cite{raghu2017svcca}, PWCCA \cite{morcos2018insights}, CKA \cite{kornblith2019similarity}, linear stitching approaches \cite{csiszarik2021similarity}, and broader surveys of representational and functional similarity measures \cite{klabunde2025similarity}.  These studies typically ask whether activations induced by the same inputs have similar geometry across networks.  Our approach is complementary: instead of comparing representations produced by the network, we compare the weight structures of the networks themselves after resolving hidden-neuron permutations.  Thus, the present work shifts the question of convergence from activation space to parameter space.

\paragraph{Platonic representations and privileged solution structures.}
Recent discussions of the Platonic Representation Hypothesis propose that sufficiently capable models trained on different data, modalities, or objectives may converge toward increasingly similar representations of the underlying world \cite{huh2024platonic}.  Our work is related to this idea, but operates at a lower level of description.  We do not ask whether different modalities converge toward a shared semantic representation.  Instead, we ask whether repeated training on the same task produces convergent organization in the trainable parameters of the system.  This distinction is important because a parameter-space attractor is not identical to a representation-space attractor.  It concerns the organization of the mechanism that computes a solution, not only the geometry of the latent variables or activations produced by that mechanism.

\paragraph{Permutation symmetries and model alignment.}
The neuron permutation problem has also become central in studies of neural-network loss landscapes and model merging.  Entezari et al. argued that many apparent barriers between independently trained networks may disappear once hidden-unit permutation symmetries are taken into account \cite{entezari2022permutation}.  Ainsworth et al. developed Git Re-Basin, explicitly aligning models modulo permutation symmetries before merging them in weight space \cite{ainsworth2023git}.  Related work on model soups shows that weight averaging can be surprisingly effective when models lie in compatible low-loss regions \cite{wortsman2022modelSoups}.  Our work shares the premise that raw weight coordinates are misleading unless permutation symmetries are handled.  However, our goal is not primarily to merge models or demonstrate low-loss interpolation.  Instead, we use alignment as a measurement tool for quantifying task-dependent structural convergence and learning dynamics.

\paragraph{Loss landscapes, connected minima, and attractor geometry.}
Several studies have challenged the naive picture of neural-network training as convergence to isolated minima.  Mode-connectivity results show that independently trained solutions can often be connected by low-loss paths \cite{garipov2018loss,draxler2018barriers}.  These findings are compatible with the idea that good solutions form extended manifolds or basins rather than isolated points.  Our results add a complementary observation: after matching hidden neurons, networks trained on the same task are not only functionally successful but also structurally closer than networks trained on different tasks, especially under metrics that emphasize important weights.  Thus, the relevant geometry appears to have both a loss-landscape aspect and a task-structural aspect.  The algorithmic-attractor interpretation attempts to connect these two views.

\paragraph{Late-stage structure and early-stage calibration.}
The observation that training can produce simple, highly organized structures is also related to neural collapse, where class means and final-layer classifiers approach a symmetric configuration during the terminal phase of classification training \cite{papyan2020neuralCollapse}.  Our study differs in two respects.  First, we examine convergence among independently trained networks rather than the internal class geometry of one network.  Second, our most surprising dynamical result concerns the opposite temporal regime: the very beginning of training.  We find that loss and accuracy improve strongly while weight changes remain extremely small and the global weight distribution is nearly unchanged.  This early phase is therefore not well described as a terminal collapse or a large morphological reorganization.  It appears instead as a highly effective calibration of structures already present in the random initialization, followed by slower structural consolidation.

\paragraph{Position of the present work.}
Taken together, previous studies show that neural networks can share activation-space structure, can be aligned modulo symmetries, and can occupy unexpectedly connected regions of low loss.  The present work combines these themes but asks a different question: whether a task leaves a reproducible structural signature directly in the matched weight matrices of independently trained networks.  The answer, in our small controlled setting, is positive.  This suggests that parameter-space analysis can contribute to the broader study of convergent learning and may provide a useful bridge between representation-level convergence, loss-landscape geometry, and the hypothesis of algorithmic attractors.

\section{Additional Information}

\subsection{Author contributions}

PK conceived and supervised the study, discussed the results, acquired funding and wrote the paper. CM supervised the study and wrote the paper. HS and NR implemented the methods and evaluated the data. AS discussed the results and acquired funding. AM and TK discussed the results and provided resources.

\subsection{Funding}
This work was funded by the Deutsche Forschungsgemeinschaft (DFG, German Research Foundation): grants KR\,5148/3-1 (project number 510395418), KR\,5148/5-1 (project number 542747151), KR\,5148/10-1 (project number 563909707) and GRK\,2839 (project number 468527017) to PK, and grants SCHI\,1482/3-1 (project number 451810794) and SCHI\,1482/6-1 (project number 563909707) to AS.

\subsection{Competing interests statement}
The authors declare no competing interests.

\subsection{Data availability statement} 
The complete data and analysis programs will be made available upon reasonable request.

\subsection{Third party rights}
All material used in the paper are the intellectual property of the authors.


\newpage
\bibliographystyle{unsrt}
\bibliography{references}


\clearpage

\clearpage

\clearpage

\clearpage

\clearpage

\clearpage

\end{document}